# Silicon nanowire band gap modification

*Michael Nolan, Sean O'Callaghan, Giorgos Fagas[*], and James C. Greer*

Tyndall National Institute, Lee Maltings, Prospect Row, Cork, Ireland

*Thomas Frauenheim*

BCCMS, Otto-Hahn-Allee, Universität Bremen, 28359 Bremen, Germany

Georgios.Fagas@tyndall.ie

**RECEIVED DATE (to be automatically inserted after your manuscript is accepted if required according to the journal that you are submitting your paper to)**

Running head title: Silicon nanowire band gap modification

.

---

[*] Author to whom correspondence should be addressed: Georgios.Fagas@tyndall.ie

Abstract

Band gap modification for small-diameter (~1 nanometer) silicon nanowires resulting from the use of different species for surface termination is investigated by density functional theory calculations. Due to quantum confinement, small diameter wires exhibit a direct band gap that increases as the wire diameter narrows, irrespective of surface termination. This effect has been observed in previous experimental and theoretical studies for hydrogenated wires. For a fixed cross-section, the functional group used to saturate the silicon surface significantly modifies the band gap resulting in relative energy shifts of up to an electron Volt. The band gap shifts are traced to details of the hybridization between the silicon valence band and the frontier orbitals of the terminating group, which is in competition with quantum confinement.

Silicon nanowires offer an alternative route for the fabrication of "end-of-the-roadmap" transistor technologies. Porous and nanocrystalline silicon has been shown to photoluminescence (1, 2) suggesting the use of silicon as a low-cost material for photonic device technologies, with the potential advantage of enabling integration of photonic devices with conventional microelectronic technologies. Recently, the reproducible synthesis of single-crystal silicon nanowires (SiNWs) with diameters as small as one nanometer has been demonstrated (3, 4, 5). The technological potential for SiNWs has been demonstrated in field-effect (6, 7, 8) and single-electron (9) transistor geometries, as well as in configurations allowing sensing of biomolecules (10, 11). The tuning of the emission wavelength in arrays of SiNWs (12) points to a possible route towards light-emitting diodes in the visible and ultraviolet regime in a fashion similar to that achieved using other Si nanostructures (13, 14).

Further applications can be explored as control over electronic and photonic properties of SiNWs is achieved via, for example, p- and n- type doping, control of wire morphology, crystal orientation, and/or modification of electronic structure. Dopant effects in nanowires have been studied experimentally (15, 16, 17) and are only recently beginning to receive theoretical attention (18, 19). There are many previous theoretical studies of Si nanowires, which have considered surface reconstruction effects in unsaturated wires (20, 21, 22) or analyzed the electronic structure of hydrogen passivated SiNWs with respect to different orientation along crystal axes and varying diameter or shape of the cross-section (18, 20, 23, 24, 25, 26, 27).

In this Letter, we investigate the effect of *chemical passivation* of the nanowire surfaces using different chemical functional groups and how they alter conclusions based solely on considering quantum confinement effects and, hence, the band gap of nanowires of a given diameter. We explore chemical passivation using different functional groups as an alternative approach for tuning the band gap of a nanowire; the magnitude of the band gap being determined by the interaction between the passivating group and the wire surface. The nanowire surface bonds are saturated with the following chemical functional groups: -H, -OH and -NH$_2$, and the resulting electronic structures are calculated using density functional theory (DFT). According to theoretical predictions quantum confinement in the direction transverse to the axis of hydrogenated SiNWs gives rise to a direct bandgap, which blue shifts with decreasing wire diameter (23, 24, 28). This feature, evident also in our calculations, is associated with the distinct photoluminescence spectra of strained-SiNWs arrays (12) and highly porous Si when interpreted as a network of SiNWs (1, 20, 23).

Whereas the quantum confinement effect on band gap is qualitatively similar for all termination species, in wires with a fixed diameter we show that hybridisation of the valence band with the frontier orbitals of the different passivating functional groups causes a considerable red-shift of the band gap relative to H-passivated SiNWs. Our prediction can be readily tested experimentally by scanning tunneling spectroscopy (STS) measurements on SiNWs. The experimental realisation of different surface passivation can be achieved following surface treatment with the appropriate terminating groups (4).

Our reference structures are silicon nanowires oriented along [100]. Density functional theory (DFT) as implemented in the VASP program package (29) with a plane wave basis expansion and an energy cut-off of 400 eV is used. The core-valence interaction is described by the projector augmented wave (PAW) method (30), with an [Ar] core for Si and a [He] core for O, S and N. The calculations are performed with cell dimensions normal to the wire axis chosen large enough to reduce interactions between neighbouring wires. A full relaxation of the ionic positions with no symmetry constraints has been performed on all nanowire structures such that the forces on each atom are less than 0.01 eV/Å. Monkhorst-Pack sampling with a 1x1x4 k-point grid and using the generalized gradient approximation (GGA) in the form of the Perdew-Burke-Ernzerhof (PBE) exchange-correlation functional are applied in the DFT computations.

We firstly analyse the structure of the nanowires. For hydrogen passivated wires we find that the. Si-Si distances are 2.36 Å in the core and 2.35 Å in the surface, with Si-H distances of 1.53 Å. In ref. (27), Vo et al. discussed that a structure with tilted hydrogen atoms was more stable than if the hydrogen atoms were symmetrically distributed about their silicon atoms, i.e. orthogonal to the wire surface. In the present work, we find that a structure with initially tilted hydrogen atoms as in ref. (27) relaxes back to the structure with symmetrically distributed hydrogen atoms. The reason for this is due to the isolated lines of dihydrides in our facets (figure 1), compared to the structures shown in figure 1 of ref. 27, where hydrogen-hydrogen interactions between parallel lines of dihydrides on a facet cause a tilting of the hydrogen atoms (27). For the wire oriented along [100], the results in (27) also show that band gap differences arising from different surface bonding within the same symmetry are only a few percent

For the –OH passivated wire with 10 Å diameter, the Si-O distances are 1.69 – 1.71 Å, while surface Si-Si distances are 2.35/2.36 Å (depending on how many –OH groups an Si atom is coordinated to), with core Si-Si distances of 2.37 Å; we find similar distances for $NH_2$ passivation. The larger diameter wires show no great change in the spread of surface Si-Si and core Si-Si distances compared to the 10 Å wire. Surface passivation does not modify the structure in any notable fashion.

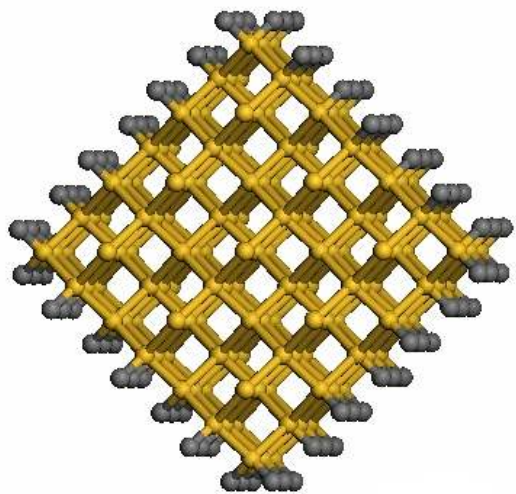

Figure 1: Structure of Si nanowire along [100]. Black spheres schematically indicate Si-bond passivation via the terminating groups X = -H, -OH, and $-NH_2$.

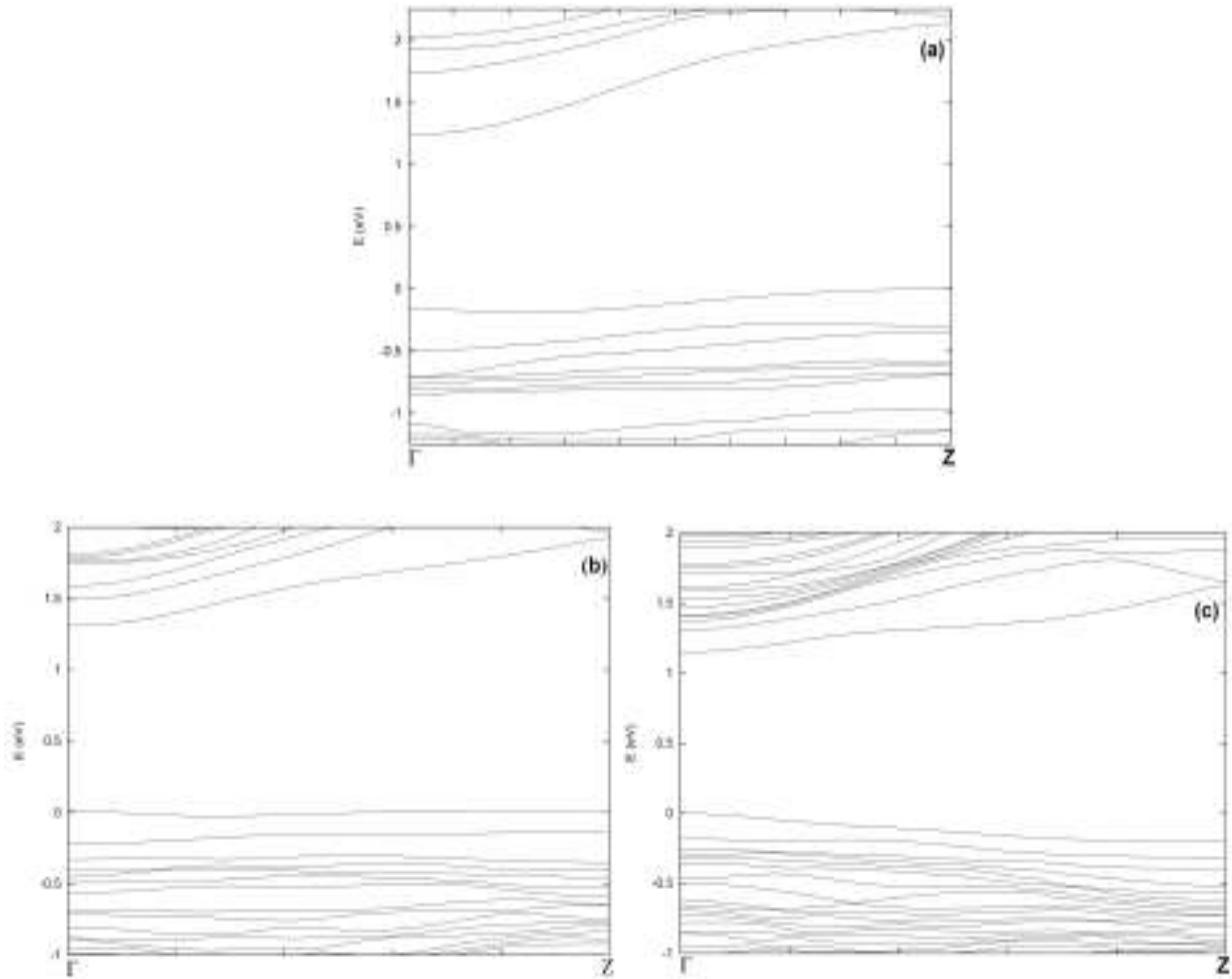

Figure 2: One-dimensional band structure for [100]-oriented silicon nanowires with -OH termination. The silicon core diameter increases from (a) to (c) (10, 14 and 17 Å, respectively). Energies are referenced with respect to the valence band edge.

Band structures for -OH terminated wires oriented along [100] are shown in Fig. 2; these may be compared to the corresponding energy dispersions reported for H-terminated SiNWs in refs. (18, 20, 23, 24, 25, 27) In Fig. 3(a), we plot the calculated band gap $E_g$ for nanowires with different surface terminations as the diameter of the silicon core changes. A direct band gap is seen in all calculations excluding the smallest diameter -OH terminated wire (31). As expected, fig. 3(a) demonstrates that the band gap decreases towards its asymptotic bulk value with increasing cross-sectional diameter. Both features of the band structures are in agreement with previous studies and can be readily explained by a zone-folding argument of the bulk Si bands onto the one-dimensional Brillouin zone of the wire (23, 28). However, in the present calculations we find that changing the surface terminating groups strongly modifies $E_g$ for a wire with fixed core diameter; there is a pronounced decrease in the magnitude of the band gap of non-hydrogen passivated wires compared to those with hydrogenated surfaces.

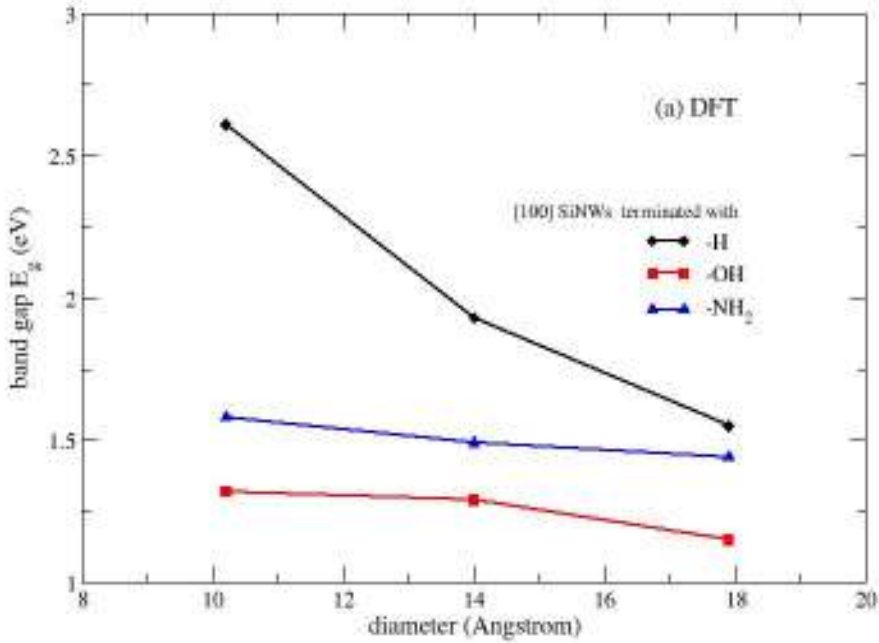

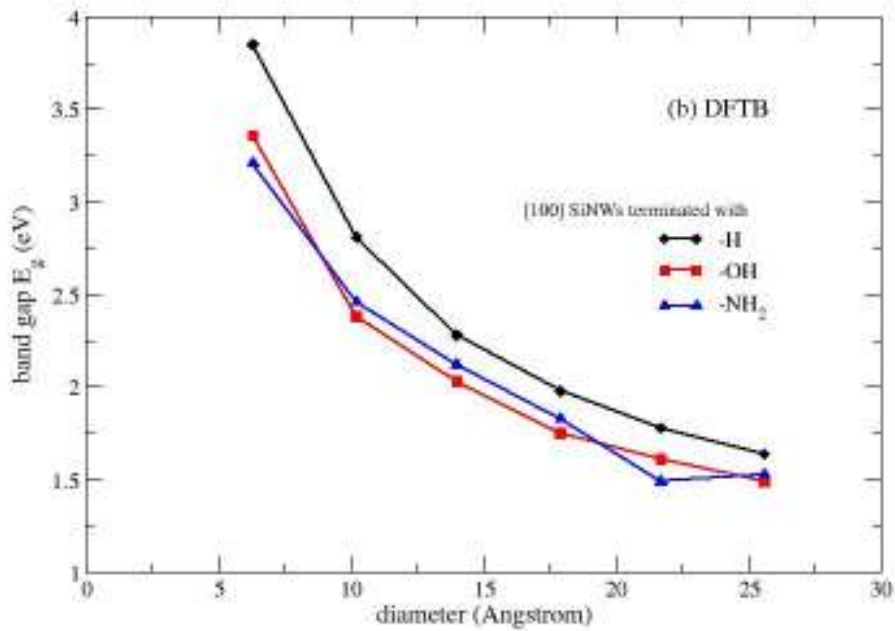

Figure 3: Band gap as a function of the [100] silicon nanowire diameter for various surface terminations. (a) DFT calculations within GGA-PBE. (b) Results from a density-functional tight-binding (DFTB) parameterization. Note the difference in scales between (a) and (b)

Hydroxyl (-OH) termination results in the narrowest band gap, followed by a relative blue-shift when surface silicon atoms are saturated by -$NH_2$ groups, with the largest band gap resulting when the surface is saturated by hydrogen. The band gap differences become larger with increasing surface-to-volume ratio pointing to a surface-induced effect. For the narrowest wire diameter, the difference of band gap energies between hydrogen-passivated wires and –OH (-$NH_2$) terminated wires is 1.3 eV (1.04 eV). The calculations demonstrate that the band gap energies in silicon nanowires can be tailored not only by the appropriate choice of diameter but also choice of surface termination. This also suggests that the band gap can be tuned by mixing the relative populations of different terminating groups on the wire, e.g. –H and –OH, and varying the percent coverage from each species. We note that similar surface chemistry effects have been previously reported by Puzder *et al.* (32) for other Si nanostructures. In their study, a single silicon site of 35-atom hydrogenated Si clusters ($Si_{35}H_{36}$) is bonded to different passivating groups. A significant gap reduction is found for double-bonded terminations while single-bonded groups, such as –OH, have minimal influence. In the present calculations, however, all hydrogen atoms from the H-passivated wire are substituted with single-bonded groups (–OH or –NH$_2$). This apparently leads to an overall stronger Si-O interaction with a correspondingly larger effect on the reduction of the band gap. This trend is also seen in the study of Si nanoclusters by Puzder *et al.* (33) when oxygen contaminates more than one surface sites; the DFT gap within LDA reduces with increasing number of oxygen substitutes (33). Both our calculations and those of ref. (32) predict a smaller band gap with –OH passivation.

DFT calculations become computationally demanding with increasing wire diameter, and for examination of larger diameter nanowire structures, at lower computational cost, we have also applied the density functional derived tight binding method (DFTB, ref.34). DFTB is based on the Harris and Foulkes formulation of DFT that expresses the energy functional around a reference atomic ground-state density. As opposed to semi-empirical tight-binding (TB), DFTB has a minimal number of input parameters and provides transferability while retaining the efficiency of TB approaches. For the present calculations a minimal basis set within GGA was used. The DFTB predicted band gap variations for the various surface terminations are plotted in fig. 3(b) and are consistent with our DFT computations. Similar results are obtained for SiNWs along [110]. Focusing on hydrogenated [110] wires, theoretical and experimental results are summarized in fig. 4. Band gap magnitudes for hydrogenated [110] wires are in reasonable agreement with the measurements of ref. (4) compared to predictions from LDA-DFT calculations (24), but this better agreement must be considered as fortuitous as DFTB should not be interpreted as a systematic approach to predicting band gaps due to the minimal basis set. However, both DFT and DFTB predict the same trends as experiment, and DFT reliably predicts the change in band gap as a function of wire diameter.

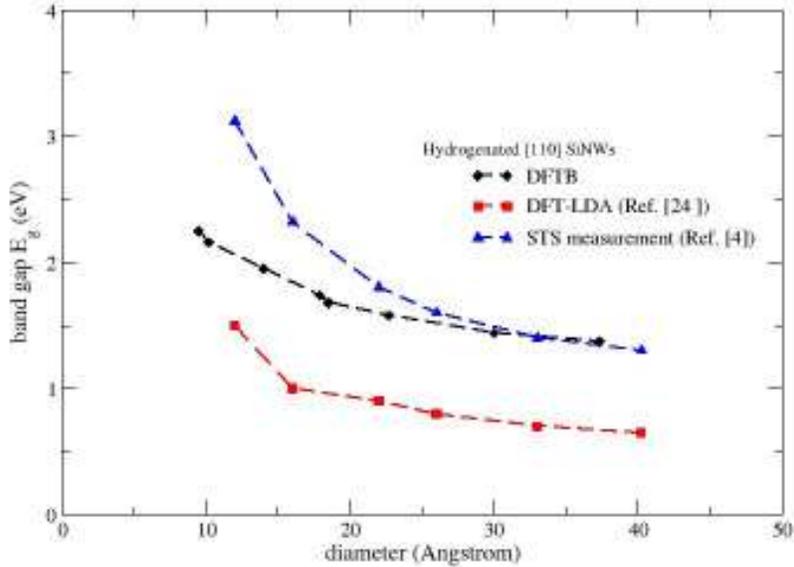

Figure 4: Band gap as a function of the [110] silicon nanowire diameter with hydrogen passivation. DFT and DFTB theoretical predictions are compared to experiment.

DFT, using approximate exchange-correlation functionals, results in a well-known and physically understood underestimation of the band gap. Consistent with this systematic underestimation of band gaps, the calculated band gap for hydrogenated [110] nanowires within a previous LDA calculation (24) is approximately 1eV smaller compared to the measured band gap using STS (4); various computational approaches can be used to correct the shortcomings of approximate DFT theories, the more accurate of these such as GW corrections are significantly computationally demanding, especially for the larger diameter wires considered in this paper. Nevertheless, DFT predictions have proven useful for prediction of trends as shown by the numerous studies of the band gap of hydrogen passivated nanowires [18, 20, 23, 24, 25, 26, 27]. For example fig. 4 shows how, although LDA underestimates the band gap of [110] nanowires, the *change* in band gap with diameter is correctly described; GW corrections shift the gap upwards to approximately the correct energy. Moreover, in ref. (32), it is shown that the trends in the band gap from LDA for silicon nanoclusters with wire diameter and/or different surface chemistry are consistent with quantum Monte Carlo calculations, giving confidence in the use of DFT for predicting *trends* in the band gap (see also ref. 33).

It is clear that due to reduction in quantum confinement with increasing wire diameter, the band gap decreases toward the bulk silicon value. However, from figs. 3(a) and 3(b), the effect for other than hydrogen-terminated wires is not as strong as observed for hydrogen saturation. This suggests a competition between the confinement effect and the role of the terminating surface groups, which has not been observed before. The atomic and angular momentum decomposed partial electronic density of states (PDOS) is used to analyze interactions between the terminating groups and SiNW band structure as displayed in fig. 5. Specifically, we decompose the total electronic density of states (DOS) into partial atomic contributions deriving either from the silicon (wire) or terminating atoms (H, O, N); the PDOS is scaled by $N_{Si} / (N_{Si} + N_X)$ for the wire and $N_X / (N_X + N_{Si})$ for the surface termination, where $N_{Si}$ and $N_X$ are the number of silicon wire atoms and terminating groups, respectively. The energy zero is set to the conduction band (CB) edge of the nanowires. H-passivated wires show no hydrogen-derived states in the valence band. For –OH and –$NH_2$ terminated wires, there is a significant weight of 2p-derived states from the oxygen and nitrogen atoms near at the valence band edge $E_v$. These states hybridize with the surface silicon 3p-states pushing $E_v$ higher in energy and forcing a reduction in the band gap. A similar effect pushes the bottom of the conduction band $E_c$ down but to a lesser extent. Partial DOS contributions are shown in Fig. 5 for wires with diameter 14 Å. The energy scale is chosen to align CB minima and the effect of hybridisation on the valence band edge is clearly seen: large traces of oxygen and nitrogen states for –OH and –$NH_2$ terminated wires are observed compared to the vanishing hydrogen contribution for –H terminated wire.

As the surface-to-volume ratio decreases with increasing diameter, the PDOS contribution of surface silicon and terminating atoms to the total DOS diminishes with eventual convergence of the band gap to its bulk value. Our results highlight how the interaction of the surface termination with the valence edge competes against the quantum confinement effect. Hence the reduction in band gap with increasing wire diameter for –OH and –NH$_2$ is less than that for H-termination where there is no interaction with the valence edge, and only confinement effects dominate. This is indeed observed in Fig. 3, where the band gap decrease with wire diameter is most notable for hydrogen termination.

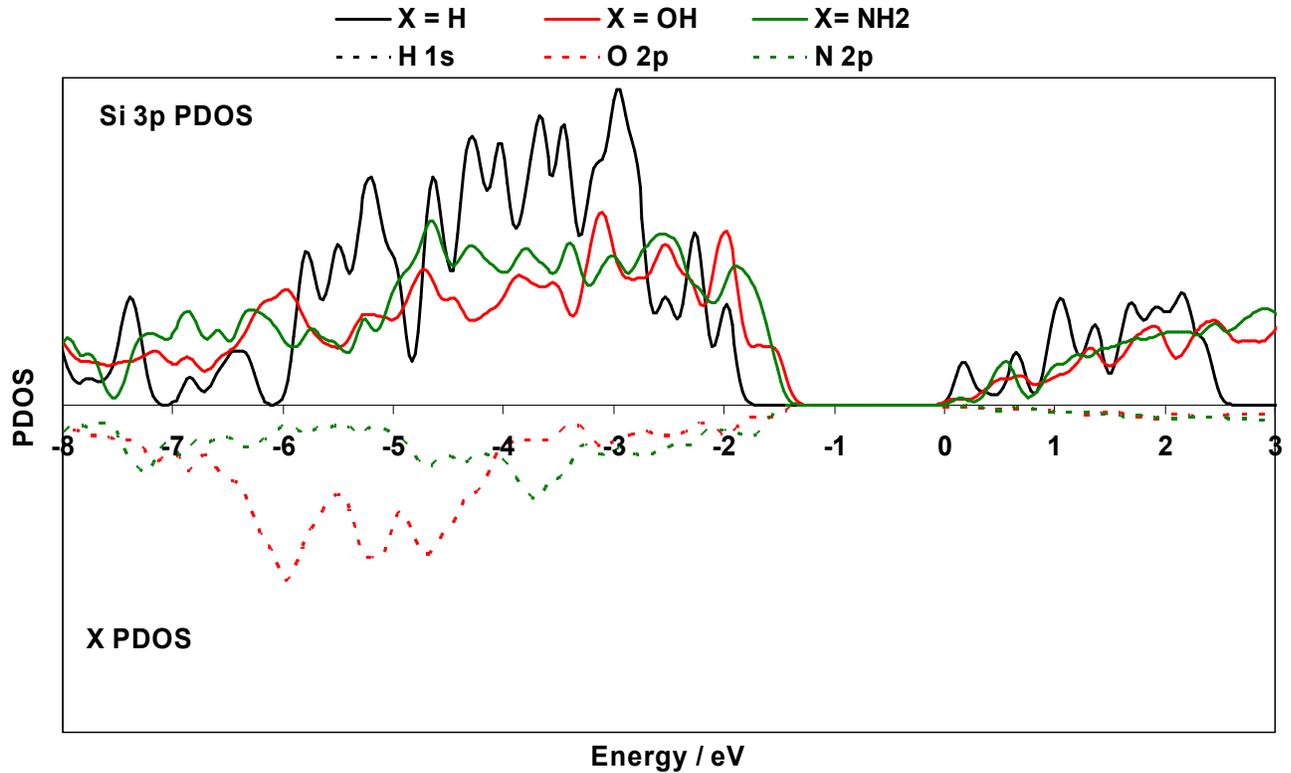

Figure 5: Partial density of states (PDOS). Electronic density of states for 14 Å diameter wires along [100] with different surface terminations X (= -H, -OH, -NH$_2$) decomposed into orbital contributions of Si, H, O, and N as indicated. Energies are referenced with respect to the conduction band edge.

In summary, we have used methods derived from first-principles to examine the effect of different surface preparation, namely, passivation of silicon surface bonds by -H, -OH and –NH$_2$, on the band gap of silicon nanowires with varying diameters. For both [100] and [110] oriented nanowires we find, in agreement with other studies, that the band gap narrows with increased wire diameter to the bulk value for large diameters. However, surface termination by –OH and –NH$_2$ introduces a hybridization effect that competes with quantum confinement at smaller diameters, inducing a large relative red shift to the band gap by up to 1 eV for small diameter wires. The origin of the reduced band gap is traced to the interaction between the Si 3p and O/N 2p states in the valence band edge of the nanowire. Our predictions can be readily verified experimentally allowing for simple electronic structure engineering of nanometer scale silicon wires via surface treatment.

**Acknowledgements**

We are grateful to the Science Foundation Ireland and the European Commission (NATCO) for supporting this work. We would also wish to acknowledge the SFI/HEA Irish Centre for High-End Computing for the provision of computational facilities.

31. In this case a slight indirect gap is observed whose origin may be linked to an oxygen-derived band. However, this feature is not observed in our DFTB calculations and deserves more elaborate investigations.